\documentclass[12pt]{article}%
\usepackage{graphicx}
\usepackage{amssymb}
\usepackage{amsmath}
\usepackage{dsfont}
\usepackage{cite}
\newcommand{\norm}[1]{\left\Vert#1\right\Vert}

\numberwithin{equation}{section}

\def\BR{{\mathbb R}}

\def\cll{{\mathcal L}}

\def\gr{{\rm grad}}
\def\si{{\sigma}}
\def\b{{\beta}}
\def\la{{\lambda}}
\def\wh{\widehat}
\def\sgn{\rm sgn}
\def\diag{\rm diag}
\def\col{\rm col}
\def\nn{\nonumber}
\newtheorem{Pa}{Paper}[section]
\newtheorem{Tm}[Pa]{{\bf Theorem}}
\newtheorem{La}[Pa]{{\bf Lemma}}
\newtheorem{Cy}[Pa]{{\bf Corollary}}
\newtheorem{Rk}[Pa]{{\bf Remark}}

\newtheorem{Dn}[Pa]{{\bf Definition}}
\newtheorem{Pn}[Pa]{{\bf Proposition}}

\title{Lippmann-Schwinger equation \\ and the connection \\ between  the scattering operator \\ and  the scattering amplitude \\ in the relativistic case}

\author{Lev Sakhnovich}

\date{}

\begin{document}

\maketitle


\begin{abstract}
In this paper, we consider two types of   the  scattering problems (relativistic case), namely,
the stationary scattering problem, where the distance $r$ tends to infinity, and
the dynamical scattering problem, where the time  $t$ tends to infinity. Using our results
on Lippmann-Schwinger
equation in the relativistic case,
we found the connection between the stationary scattering problem (the scattering amplitude) and
the dynamical scattering problem (the scattering operator). This result  is the quantum mechanical
analog of  the ergodic formulas in the classical  mechanics.\end{abstract}
\section{Introduction}
This work is a continuation and application of our work \cite{Sakh1} on Lippmann-Schwinger equation in the 
relativistic case. 
The classical integral Lippmann--Schwinger equation plays an essential role in the non-relativistic scattering theory \cite{IK}.
Among interesting recent references one could mention, for instance,  \cite{AGeL, Co, Eg, Ge, Ko, Ro} (and many further references may be found therein).
The relativistic analogue of the Lippmann--Schwinger equation was formulated in  terms of the limit
values of the corresponding resolvent \cite{Bogo, Land}. In  \cite{Sakh1} we found the limit values of this resolvent
in the explicit form, which opened way to interesting further developments. Some of these developments are given in \cite{Sakh2},
and  the present paper is dedicated to new important results in this direction.

Consider the Dirac operator
\begin{equation}i\frac{\partial}{\partial{t}}u(r,t)=\mathcal{L}u(r,t),\label{1.1}\end{equation}
where $u(r,t)$ is a $4\times1$ vector function, $r=(r_1,r_2,r_3)$, and the operators $\mathcal{L}$ and $\mathcal{L}_{0}$ are defined by the relations
\begin{equation}\mathcal{L}u=[-e\nu(r)I_{4}+m\beta+{\alpha}({p}+e{A}(r))]u,\quad
\mathcal{L}_{0}u=(m\beta+{\alpha}{p})u. \label{1.2}\end{equation}
Here $I_k$ is the $k\times k$ identity matrix, $p=-i \, \gr$, $\nu$ is a scalar potential, $A$ is a vector potential and $(-e)$ is the electron charge.
Next we define $\alpha=[\alpha_1,\alpha_2,\alpha_3]$ and $\b$. The matrices $\alpha_s$  and $\b$ are the $4{\times}4$ matrices of the forms
\begin{equation}\alpha_s=\left(
                           \begin{array}{cc}
                             0 & \sigma_s \\
                             \sigma_s & 0 \\
                           \end{array}
                         \right), \quad s=1,2,3; \quad 
                          \beta=\left(
                         \begin{array}{cc}
                            I_2 & 0 \\
                           0 & -I_2 \\
                         \end{array}
                       \right),
                         \label{1.3}\end{equation}
where Pauli matrices $\si_s$ are given by the formulas
\begin{equation}\sigma_1=\left(
                           \begin{array}{cc}
                             0 & 1 \\
                             1 & 0 \\
                           \end{array}
                         \right),\quad
\sigma_2=\left(
                           \begin{array}{cc}
                             0 & -i \\
                             i & 0 \\
                           \end{array}
                         \right),\quad
\sigma_3=\left(
                           \begin{array}{cc}
                             1 & 0 \\
                             0 & -1 \\
                           \end{array}
                         \right).\label{1.4}\end{equation}

In this paper, we consider two types of   the  scattering problems (relativistic case), namely,
the stationary scattering problem, where the distance $r$ tends to infinity, and
the dynamical scattering problem, where the time  $t$ tends to infinity. 
In Section 2, we formulate our results on relativistic  Lippmann-Schwinger equation \cite{Sakh1}, and   our approach to the scattering problems is essentially based on these results.
In Section 3, we study the mentioned above dynamical scattering problem and the corresponding
scattering operator. We present the conditions under which the wave operators exist and are complete. Hence, the corresponding scattering operator 
$S(\mathcal{L},\mathcal{L}_{0})$ also exists, and it is unitary.
We construct  the energetic representation of $S(\mathcal{L},\mathcal{L}_{0})$.
In Section 4, we consider the  scattering amplitude $f(\omega,\omega^{\prime},n,\lambda)$ and describe the interconnections between the   scattering amplitude
and the scattering operator  $S(\mathcal{L},\mathcal{L}_{0})$. Finally, in Section 5 we  study separately the case where the potential $V(r)$ satisfies the condition:
$V(r)=V(|r|)$.
\begin{Rk}\label{Remark 1.1}The ergodic results  for  Schr\"{o}dinger equation have been obtained earlier in
\cite{Sakh6}.\end{Rk}
\section{Preliminaries}
1. \emph{Let us introduce a relativistic analog of the Lippmann-Schwinger equation
$($RLS equation$)$  constructed in  \cite{Sakh1}}. First, put
\begin{equation}B_{+}(r,\lambda)=Q(r)+(2\pi)^{3/2}\lambda^{2}Q(r)\ast{J_{+}(r,\lambda)}+
{\lambda}J_{+}(r,\lambda),\label{ 2.1}\end{equation}
where  $\lambda=\overline{\lambda},\, |\lambda|>m$, $\varkappa=\sqrt{\lambda^2-m^2}$,
\begin{equation}J_{+}(r,\lambda)=-\sqrt{\frac{\pi}{2}}\exp\{{}i\varkappa|r|\}/|r|,\quad \lambda>m,\label{2.2}\end{equation}
\begin{equation}J_{+}(r,\lambda)=-\sqrt{\frac{\pi}{2}}\exp\{{-}i\varkappa|r|\}/|r|,\quad \lambda<-m,\label{2.3}\end{equation}
and the matrix function $Q(r)$ has the form
\begin{equation} Q(r)=\sqrt{\pi/2}\exp\{-m|r|\}[m\beta+i(m+1/|r|)r\alpha/|r|]/|r|.\label{2.4}\end{equation}
Here, $r\alpha=r_1\alpha_1+r_2\alpha_2+r_3\alpha_3$, the  matrices $\alpha_k$ and $\beta$ are defined by the relations \eqref{1.3} and \eqref{1.4},
and $\ast$ denotes the convolution:
\begin{equation} F(r)\ast{G(r)}=\int_{\BR^3}F(r-v)G(v)dv. \label{2.5}\end{equation} 
The RLS equation takes the form  \cite{Sakh1}:
\begin{equation}\phi(r,k,n)=\exp\{ik{\cdot}r\}{\wh g}_{n}(k)-(2\pi)^{-3/2}\int_{\BR^3}B_{+}(r-s,\lambda)V(s)\phi(s,k,n)ds,
\label{2.6}\end{equation} 
where $k=(k_1,k_2,k_3)$ are vectors from $\BR^3$ such that $|k|=\varkappa=\sqrt{\lambda^2-m^2}$, 
\begin{align}
&
V(r):=-e{\nu}(r)I_4+e{\alpha}A(r),\label{2.7}
\end{align}
the vectors $\wh {g}_{n}(k)$ have the form $\wh {g}_{n}(k)=g_{n}(k)/|g_{n}(k)|$ and $g_{n}(k)$ are given by 
\begin{align}&
g_1=\begin{bmatrix}(-k_1+ik_2)/(m+\lambda_3) \\ k_3/(m+\lambda_3) \\ 0 \\ 1\end{bmatrix}, \quad
g_2=\begin{bmatrix}-k_3/(m+\lambda_3) \\ (-k_1- ik_2)/(m+\lambda_3) \\1 \\ 0 \end{bmatrix},
\label{2.8}
\\ &
g_3=\begin{bmatrix} (-k_1+ik_2)/(m-\lambda_3)\\ k_3/(m-\lambda_3)\\ 0 \\ 1\end{bmatrix},
\quad
g_4= \begin{bmatrix} -k_3/(m-\lambda_3) \\ (-k_1-ik_2)/(m-\lambda_3) \\ 1 \\ 0\end{bmatrix};
\label{2.9}
\\ &\nn
\lambda_n=-\sqrt{k^2+m^2} \quad {\mathrm{for}} \quad
n=1,2;  \quad \lambda_n=\sqrt{k^2+m^2} \quad {\mathrm{for}} \quad
n=3,4,
\end{align}
and the positive branch of square root is chosen.
\begin{Rk} Clearly $\phi$ above depends on $\la$ but we omit sometimes some
variables in the text for convenience.
\end{Rk}
Under some natural 
conditions, $\phi$ is a solution (in a distributive sense) of the equation $\cll\phi=\la \phi$
(see Theorem \ref{Theorem 2.7}).

2. Further we assume that the matrix $V(r)$ is self-adjoint:
\begin{equation}V(r)=V^*(r).\label{2.10}\end{equation}
 Hence, $V(r)$ can be represented in the form
\begin{equation}V(r)=U(r)D(r)U^{*}(r),\label{2.11}\end{equation}
where $U(r)$ is an unitary matrix and $D(r)$ is a diagonal matrix:
\begin{equation}D(r)=\diag\{d_{1}(r),d_{2}(r),d_{3}(r),d_{4}(r)\}.\label{2.12}\end{equation}
Let us introduce the diagonal matrices
\begin{align} & D_{1}(r):=\diag\{|d_{1}(r)|^{1/2},|d_{2}(r)|^{1/2},|d_{3}(r)|^{1/2},|d_{4}(r)|^{1/2}\},
\label{2.13}
\\ &
W(r):=\diag\{\sgn \, d_{1}(r),\sgn \, d_{2}(r),\sgn \, d_{3}(r),\sgn \, d_{4}(r)\}.\label{2.14}\end{align}
Formulas \eqref{2.11}-\eqref{2.14} imply that $V$ admits representation
\begin{equation}V(r)=V_{1}(r)W_{1}(r)V_{1}(r),\label{2.15}\end{equation}
where
\begin{equation}V_{1}(r):=U(r)D_{1}(r)U^{*}(r),\quad W_{1}(r):=U(r)W(r)U^{*}(r).\label{2.16}\end{equation}
It is easy to see that
\begin{equation}\|V_{1}(r)\|^{2}=\|V(r)\|,\quad \|W_{1}(r)\|=1.\label{2.17}\end{equation}

We will need also a \emph{modified RLS integral equation.}
If the $4{\times}1$ matrix function $\phi(r,k,n)$ is a solution of the RLS equation \eqref{2.6}, then the matrix function $\psi(r,k,n)=V_{1}(r)\phi(r,k,n)$
is a solution of the following  modified RLS integral equation:
\begin{equation}\psi(r,k,n)=\exp\{ik{\cdot}r\}V_{1}(r)\wh {g}_{n}(k)-(2\pi)^{-3/2}B_{+}(\lambda)\psi(r,k,n),
\label{2.18}\end{equation}
where
\begin{equation}B_{+}(\lambda)f=\int_{\BR^3}V_{1}(r)B_{+}(r-s,\lambda)V_{1}(s)W_{1}(s)f(s)ds.
\label{2.19}\end{equation} We note that the operators $B_{+}(\lambda)$ act   in the Hilbert space
$L^{2}(\BR^3)$. (We say that the matrix belongs to the Hilbert space
$L^{2}(\BR^3)$ if each element of the matrix belongs to the Hilbert space
$L^{2}(\BR^3)$.)
In  \cite{Sakh1}, we proved the following result.
\begin{Tm}\label{Theorem 2.1} If
the function  $\|V(r)\|$ is bounded and belongs to the space $L^{1}(\BR^3)$
then the operator $B_{+}(\lambda)$ is compact.
\end{Tm}
3. Let us introduce the following definition.
\begin{Dn}\label{Definition 2.2} \cite{Sakh1}. We say that $\lambda {\in}E=(-\infty,-m]\cup[m,+\infty)$ is an exceptional value if the equation 
$$[I+(2\pi)^{-3/2}B_{+}(\lambda)]\psi=0$$ 
has a nontrivial solution in the space $L^2(\BR^3)$. The set of exceptional points is denoted by $\mathcal{E}_{+}$.

 The set
 of  points $\lambda$ such that $\lambda{\in}E$ and  $\lambda{\notin}\mathcal{E}_{+}$ is denoted by $E_{+}$.
 \end{Dn}
 We have \cite{Sakh1}:
\begin{La}\label{Lemma 2.3} Let the conditions of Theorem \ref{Theorem 2.1} be fulfilled.  Then, equation \eqref{2.18} has one and only one
solution $\psi(r,k,n)$ in $L^2(\BR^3)$ for each $\la \in E_{+}$. \end{La}
\begin{Cy}\label{Corollary 2.4} Let the conditions of Theorem \ref{Theorem 2.1} be fulfilled. 
Then, equation \eqref{2.6} has one and only one
solution $\phi(r,k,n)$ which satisfies the condition $V_{1}(r)\phi(r,k,n){\in}L^2(\BR^3)$  for each $\la \in E_{+}$. \end{Cy}
4. The interconnections between spectral and scattering results are described in \cite{Sakh1} by the following theorem.
\begin{Tm}\label{Theorem 2.5} Assume that $V(r)=V^*(r)$ and the function $\|V(r)\|$ is bounded and belongs to the space $L^{1}(\BR^3)$. Let $\lambda{\in}E_{+}$
and the relation
\begin{equation}\|V_1(r)\|=O(|r|^{-3/2}),\quad |r|{\to}\infty,\label{2.20}\end{equation}
hold. Then, 
the  asymptotics of the solution $\phi(r,k,n,\lambda)$ of the RLS equation \eqref{2.6} is given by the formula
\begin{equation}\phi(r,k,n,\lambda)=\exp\{ik{\cdot}r\}\wh {g}_{n}(k)+\big(\exp\{i\varkappa|r|\}/{|r|}\big)
f(\omega,\omega^{\prime},n,\lambda)
+o(1/|r|) \label{2.21}\end{equation}
where  $|r|{\to}\infty$, $\omega=r/|r|,\, \omega^{\prime}=k/|k|$, and
\begin{equation}f(\omega,\omega^{\prime},n,\lambda)=\frac{\lambda}{4\pi}\int_{\BR^3}\exp\{-i{\varkappa}s{\cdot}\omega\}
V(s)\phi(s,k,n,\lambda)ds.  \label{2.22}\end{equation}
 \end{Tm}
(Recall that $k$ is a vector from the space $\BR^3$  and $\varkappa=|k|=\sqrt{\lambda^2-m^2}.$)
 \begin{Dn}\label{Definition 2.6} The $4{\times}1$ vector functions $f(\omega,\omega^{\prime},n,\lambda)$ given by \eqref{2.22} are called the relativistic scattering
amplitudes.\end{Dn}
5. Now we will formulate the connections between the solutions of the equation
\begin{equation}\mathcal{L}\phi=\lambda\phi\label{2.23}\end{equation}
and the solutions of the  RLS  equation \eqref{2.6} (see  \cite{Sakh1}).
As usual, the solution $\phi$ of \eqref{2.23}, which belongs to $L^{2}(\BR^3)$, is called the eigenfunction of $\cll$.
\begin{Tm}\label{Theorem 2.7}  Let the conditions of Theorem \ref{Theorem 2.5} be fulfilled and let  the function $\phi(r,k,n)$ be a solution of the 
RLS equation \eqref{2.6} such that 
$$V_{1}(r)\phi(r,k,n){\in}L^{2}(\BR^3).$$ 
Then, the vector function $\phi(r,k.n)$
is the solution of the equation \eqref{2.23}
in the distributive sense.
\end{Tm}
Recall that $\lambda{\in}E_{+}$ is required in Theorem \ref{Theorem 2.5}.
\begin{Cy}\label{Corollary 2.8} Let all the conditions of Theorem \ref{Theorem 2.5}, excluding the relation $\lambda{\in}E_{+}$, be fulfilled. If
$\lambda{\in}E$ is an eigenvalue of the corresponding operator $\mathcal{L}$, then $\lambda{\in}\mathcal{E}_{+}$.\end{Cy}
\section{The scattering operator}
1. In this section, we use the  relativistic Lippmann-Schwinger equation
to study the  scattering problem. We introduce the operator function
\begin{equation}\Theta(t)=\exp\{it\mathcal{L}\}\exp\{-it\mathcal{L}_{0}\}.
\label{3.1}\end{equation}
The absolutely continuous subspaces with respect to the operators $\mathcal{L}$ and $\mathcal{L}_0$
are denoted by by $G$ and $G_0$,
 respectively.
The wave operators $W_{\pm}(\mathcal{L},\mathcal{L}_0)$ are defined (see \cite{RS}) by the relation:
\begin{equation}W_{\pm}(\mathcal{L},\mathcal{L}_0)=\lim_{t{\to}\pm{\infty}}\Theta(t)P_0.\label{3.2}
\end{equation} 
Here,  $P_0$ is the orthogonal projector
on the  subspace $G_0$, and the
limit in \eqref{3.2} is the limit  in the sense of strong convergence.
\begin{Rk}\label{Remark 3.1} In this paper, we consider the case of the operator $\mathcal{L}_{0}$ defined by  \eqref{1.2}. In this case, we have: $G_{0}=\BR^3$ and $P_{0}=I$.\end{Rk}
\begin{Tm}\label{Theorem 3.2} \cite{Sakh1}. Assume that $V(r)=V^{*}(r)$, and
the function  $\|V(r)\|$ is bounded, belongs to the space $L^{1}(\BR^3)$ and
\begin{equation}\int_{-\infty}^{+\infty}\left[\int_{|r|>\varepsilon |t|}\|V(r)\|^{2}dr\right]^{1/2}dt<\infty,\quad \varepsilon>0. \label{3.3}\end{equation}
Then, the wave operators $W_{\pm}(\mathcal{L},\mathcal{L}_0)$ exist.
\end{Tm}
\begin{Cy}\label{Corollary 3.3}Let the condition  $V(r)=V^{*}(r)$ be fulfilled. If
the function  $\|V(r)\|$ is bounded and
\begin{equation}\|V(r)\|{\leq}\frac{M}{|r|^{\alpha}},\quad |r|{\geq}\delta>0,\quad \alpha>3,
\label{3.4}\end{equation}
then the wave operators $W_{\pm}(\mathcal{L},\mathcal{L}_0)$ exist.
\end{Cy}
\begin{Dn}\label{Definition 3.4}The scattering operator $S(\mathcal{L},\mathcal{L}_0)$ is defined by the relation
\begin{equation}S(\mathcal{L},\mathcal{L}_0)=W_{+}^{*}(\mathcal{L},\mathcal{L}_0)W_{-}(\mathcal{L},\mathcal{L}_0).
\label{3.5}\end{equation}\end{Dn}
\begin{Dn}\label{Definition 3.5} Suppose that the wave operators $W_{\pm}(\mathcal{L},\mathcal{L}_{0})$
exist. They are complete if the images of these operators coincide with $G$.
\end{Dn}
\begin{Tm}\label{Theorem 3.6} \cite{Sakh2}. Let the conditions of   Theorem \ref{Theorem 3.2} be fulfilled. Then the corresponding wave operators $W_{\pm}(\mathcal{L},\mathcal{L}_{0})$ exist and are complete. \end{Tm}

2. Now, we will find the energetic representation $S(\lambda)$ of  the scattering  operator $S(\mathcal{L},\mathcal{L}_0).$ In order to do it, we use  Definition 3.4
and  write
 \begin{equation} (f,(S-I)g)=((W_{+}-W_{-})f,W_{-}g),\label{3.7}\end{equation}
where $(f,g)$ stands for the scalar product of $f$ and $g$. We assume that conditions of the Theorem 3.2 are fulfilled. Hence, the wave operators $W_{\pm}$ and the scattering operator $S$ exist. We assume also that $f(r)$ and $g(r)$ belong to $C_{0}^{\infty}$. Using  these assumptions we can change the order of integration in the calculations below.
It follows from  relation \eqref{3.7} and equality
\begin{equation}\frac{d}{dt}[\exp\{it\mathcal{L}\}\exp\{-it\mathcal{L}_0\}]=
i \exp\{it\mathcal{L}\}V \exp\{-it\mathcal{L}_0\},\label{3.8}\end{equation}
that
\begin{equation}
(f,(S-I)g)=i\lim_{T{\to}\infty}\int_{-T}^{T}(\exp\{it\mathcal{L}\}V\exp\{-it\mathcal{L}_0\}f,W_{-}\, g)dt.
\label{3.9}\end{equation}
Taking into account Abel's limits, we obtain  (see \cite{RS}, section 6, Lemma 5)
\begin{equation} (f,(S-I)g)=i\lim_{\delta{\to}+0}\int_{-\infty}^{\infty}\exp\{-\delta|t|\}
\big(\exp\{it\mathcal{L}\}V\exp\{-it\mathcal{L}_0\}f,W_{-}\, g\big)dt.\label{3.10}\end{equation}
Introduce the vector functions
\begin{equation}F(r,t)=\exp\{it\mathcal{L}\}V\exp\{-it\mathcal{L}_0\}f,\quad G(r)= W_{-}g,
\label{3.11}\end{equation}
and consider the integral
\begin{equation}\widetilde{F}_{n}(k,t)=(2\pi)^{-3/2}\int_{\BR^3}\phi^{*}(r,k,n,\lambda_{n}(k))F(r,t)
dr, \label{3.12}\end{equation}
where $\lambda_{n}(k)=-\sqrt{k^2+m^2},$ if $n=1,2$, and $\lambda_{n}(k)=\sqrt{k^2+m^2},$ if $n=3,4.$
It follows from \eqref{3.11} and \eqref{3.12} that
\begin{align}\widetilde{F}_{n}(k,t)=&(2\pi)^{-3/2}
\\ \nn & \times
\int_{\BR^3}\phi^{*}(r,k,n,\lambda_{n}(k))\exp\{it\lambda_{n}(k)\}
V(r)[\exp\{-it\mathcal{L}_{0}\}f(r)]dr.
 \label{3.13}\end{align}
Let us introduce the domain  $D(R,\varepsilon,k)$  such that
 $|k|<R$ and
$|k-k_{0}|>\varepsilon>0$ for all $k_{0}$ satisfying the condition
$\pm\sqrt{k_{0}^{2}+m^2}{\in}\mathcal{E}_{+}$.
 We will use the following notation:
\begin{equation} \int_{D}  \, \cdot \, dk=\lim\int_{D(R,\varepsilon,k)}  \, \cdot \, dk,\quad {\mathrm{for}} \quad  R{\to}\infty,\quad
\varepsilon{\to}+0.\label{3.14}\end{equation}
According to  \cite[(4.50)]{Sakh2}, we have
\begin{equation}(F(r,t),G(r))= \int_{D}\sum_{n=1}^{4}[\widetilde{F}^{*}_{n}(k,t)\widetilde{g}_{0n}(k)]dk,
 \label{3.15}\end{equation} where  $\widetilde{g}_{0,n}$ and $\phi_{0}$ are defined by the relations
\begin{align} & \widetilde{g}_{0,n}(k)=(2\pi)^{-3/2}
\int_{\BR^3}\phi_{0}^{*}(r,k,n)g(r)dr \quad (1{\leq}n{\leq}4),
 \label{3.16}
 \\ &
 \phi_{0}(r,k,n)=\exp\{ir{\cdot}k\}\wh {g}_{n}(k).
 \label{3.17}\end{align}
The inverse  to the transformation \eqref{3.16} has the form (see \cite{Sakh2}):
\begin{equation}f(r)=(2\pi)^{-3/2}\int_{D}\exp\{ir{\cdot}q\}\sum_{n=1}^{4}[\wh {g}_{n}(q)\widetilde{f}_{n}(q)]dq.
\label{3.18}\end{equation}
Using \eqref{3.18} we have
\begin{equation}\exp\{-it\mathcal{L}_0\}f(r)=(2\pi)^{-3/2}\int_{D}\exp\{ir{\cdot}q\}\sum_{n=1}^{4}\exp\{-i\lambda_{n}(q)t\}[\wh {g}_{n}(q)\widetilde{f}_{n}(q)]dq.
 \label{3.19}\end{equation}
We need the following notations
\begin{equation}T(q,k,p):=(2\pi)^{-3}Z_{p}^{*}(q)\int_{\BR^3}\exp\{-iq{\cdot}r\}V(r)\Phi_{p}(r,k)dr,\label{3.20}
\end{equation}
where
\begin{equation}Z_{1}(q)=[\wh {g}_{1}(q),\, \wh {g}_{2}(q)], \quad Z_{2}(q)=[\wh {g}_{3}(q),\, \wh {g}_{4}(q)],\label{3.21}\end{equation}
and
\begin{equation} \Phi_{1}(r,k)=[\phi(r,k,1),\, \phi(r,k,2)], \quad \Phi_{2}(r,k)= [\phi(r,k,3),\, \phi(r,k,4)].
\label{3.22}\end{equation}
Integrating \eqref{3.10} over the variable t and using formulas \eqref{3.19}--\eqref{3.22} we rewrite the
relation \eqref{3.10} in the form:
\begin{equation} (f,(S-I)g)=i\lim_{\delta{\to}+0}\left[\sum_{1{\leq}s,n{\leq}2}J_{\delta}(s,n,1)+
\sum_{3{\leq}s,n{\leq}4}J_{\delta}(s,n,2)\right],\label{3.23}
\end{equation}
where
\begin{align}& J_{\delta}(s,n,p) \label{3.24}
\\ \nn &
=\int_{D}\int_{\BR^3}
\frac{2\delta}{\delta^2+[\lambda_{s}(q)-\lambda_{n}(k)]^2}
\overline{[\widetilde{f}_{s}(q)]}T_{s,n}(q,k)dq[\widetilde{g}_{0,n}(k)]dk,
\\ &
T(q,k,1)=\left(
                \begin{array}{cc}
                  T_{1,1}(q,k) & T_{1,2}(q,k) \\
                  T_{2,1}(q,k) & T_{2,2}(q,k) \\
                \end{array}
              \right),\label{3.25}
\\ &
T(q,k,2)=\left(
                \begin{array}{cc}
                  T_{3,3}(q,k) & T_{3,4}(q,k) \\
                  T_{4,3}(q,k) & T_{4,4}(q,k) \\
                \end{array}
              \right).\label{3.26} \end{align}
 Let us consider the case $\lambda_{s}(k)=\lambda_{n}(k)$ in greater detail.
\begin{La}\label{Lemma 3.7} If $\lambda(|q|)=\sqrt{q^2+m^2}$, then
\begin{equation}M:=\lim_{\delta{\to}+0}\int_{0}^{\infty}
\frac{2\delta}{\delta^2+[\lambda(|q|)-\lambda(|k|)]^2}
d|q|={\pi}\sqrt{k^2+m^2}|k|^{-1}.\label{3.27}\end{equation}
\end{La}
\emph{Proof}.
We represent $\lambda(|q|)-\lambda(|k|)$ in the form
\begin{equation}\lambda(|q|)-\lambda(|k|)=\frac{q^2-k^2}{\lambda(|q|)+\lambda(|k|)}.\label{3.28}\end{equation}
If $ |q|{\sim}|k|$, the relation
\begin{equation}\lambda(|q|)-\lambda(|k|){\sim}\frac{|k|(|q|-|k|)}{\sqrt{k^2+m^2}}\label{3.29}\end{equation}
is valid.
Hence, we obtain the equality
\begin{equation}M=(k^2+m^2)\lim_{\delta{\to}+0}\int_{0}^{\infty}
\frac{2\delta}{\delta^2(k^2+m^2)+k^2(|q|-|k|)^2}
d|q|.\label{3.30}\end{equation}
If   $4ac-b^2>0$, the equality
\begin{equation}\int\frac{dx}{ax^2+bx+c} =\frac{2}{\sqrt{4ac-b^2}}\arctan\frac{2ax+b}{\sqrt{4ac-b^2}}
\label{3.31}\end{equation}
holds, and in our case (see \eqref{3.28}) we have this situation because \\
$a=k^2,\,b=-2|k|^3,\,c=k^4+\delta^2(k^2+m^2),$ that is,
\begin{equation} 4ac-b^2=4k^2\delta^2(k^2+m^2)>0.\label{3.32}\end{equation}
Relations  \eqref{3.30} and \eqref{3.32} imply
\begin{equation}M={\pi}\sqrt{k^2+m^2}|k|^{-1}.\label{3.33}\end{equation}
$\Box$

According to \eqref{3.20}, \eqref{3.25} and \eqref{3.26}, the  functions  $T_{s,n}(q,k,p)$ are bounded and continuous. Then,
using spherical system of coordinates in the space $k{\in}\BR^3$ and relation \eqref{3.27} we obtain
\begin{align}& \label{3.34} \lim_{\delta{\to}+0}J_{\delta}(s,n,p)
\\ & \nn
=\int_{D}\overline{[\widetilde{f}_{s}(|k|\omega)]}a(|k|)
\int_{S^2}T_{s,n}(|k|\omega,|k|\omega^{\prime})[\widetilde{g}_{0,n}(|k|\omega^{\prime})]d\Omega(\omega^{\prime})dk,
\end{align}
where $\omega=q/|q|,\quad \omega^{\prime}=k/|k|$ and
\begin{equation}a(|k|)={\pi}\sqrt{k^2+m^2}|k|. \label{3.35}\end{equation}
Here,  $S^2$ denotes the surface $|q|=1$ in $\BR^3$ and $d\Omega$ is the  standard measure on $S^2$.
Using  \eqref{3.34}  we construct the operators
\begin{equation}T_{p}(\lambda)h_{1}(\omega)=a(|k|)\int_{S^2}T(|k|\omega,|k|\omega^{\prime},p)
{h}(\omega^{\prime})d\Omega(\omega^{\prime}),\label{3.36}\end{equation}
where    $h(\omega)$  are  $2{\times}1$ vector functions.
Let us  introduce the Hilbert space $\mathcal{H}$ of the vector functions $h(\omega)=\col[h_{1}(\omega),h_{2}(\omega)]$, where col stands for column. 
The norm in the space $\mathcal{H}$ is defined by the relation
\begin{equation}\|h\|^{2}=\int_{S^2}(|h_{1}(\omega)|^{2}+|h_{2}(\omega)|^{2})d\Omega(\omega).\label{3.37}\end{equation}
The operators $T_{p}(\lambda)$   act in the space $\mathcal{H}.$
\begin{Pn}\label{Proposition 3.8} Let the conditions of Theorem \ref{Theorem 2.5} be fulfilled. Then, the energetic representation $S(\lambda)$ of  the scattering  operator $S(\mathcal{L},\mathcal{L}_0)$ is unitary in the space $\mathcal{H}$ and has the form
\begin{equation}S(\lambda)= S_{1}(\lambda),\, \lambda<-m; \quad S(\lambda)= S_{2}(\lambda),\, \lambda>m,
\label{3.38}\end{equation}
where
\begin{equation}S_{1}(\lambda)=I+iT_{1}(\lambda),\quad S_{2}(\lambda)=I+iT_{2}(\lambda).\label{3.39}\end{equation}
\end{Pn}
\emph{Proof.} Taking into account \cite[Proposition 4.11]{Sakh2}, we see
 that the operator $S(\lambda)$  is unitary in the space $\mathcal{H}.$  Relations \eqref{3.23},
\eqref{3.34} and \eqref{3.36} imply relations \eqref{3.38} and \eqref{3.39}. $\Box$

3. Next, we study   the structure of the operators
 $T_{p}(\lambda)$. For that purpose we introduce the operator
\begin{equation}\mathcal{F}_{p}(\lambda)f(r)=(2\pi)^{-3/2}\sqrt{a(|k|)}\int_{\BR^3}
\exp\{-iq{\cdot}r\}Z_{p}^{*}(q)V_1(r)f(r)dr,\label{3.40}\end{equation}
where $q=|k|\omega.$  The functions $Z_{p}(q)$ and  $a(|k|)$ above are defined by \eqref{3.21} and \eqref{3.35}, respectively.
The adjoint to $\mathcal{F}_{p}(\lambda)$ operator has the form
\begin{equation}\mathcal{F}^{*}_{p}(\lambda)h(\omega)=(2\pi)^{-3/2}\sqrt{a(|k|)}V_{1}(r)\int_{S^2}e^{i|k|\omega{\cdot}r}
Z_{p}(|k|\omega)h(\omega)d\Omega(\omega)
\label{3.41}\end{equation}
The following assertion is valid.
\begin{Pn}\label{Proposition 3.9} Let the conditions of Theorem \ref{Theorem 2.5} be fulfilled. If $\lambda$  belongs to
$E_{+}$ then  the operators $T_{p}(\lambda)$ can be represented in the form
\begin{equation}T_{p}(\lambda)=\mathcal{F}_{p}(\lambda)W_{1}(r)[I+(2\pi)^{-3/2}B_{+}(\lambda)]^{-1}
\mathcal{F}^{*}_{p}(\lambda)\label{3.42}\end{equation}\end{Pn}
\emph{Proof.}
 According to \eqref{2.18}, \eqref{3.21} and \eqref{3.22} we have
\begin{equation}V_{1}(r)\Phi_{p}(r,k)=[I+(2\pi)^{-3/2}B_{+}(\lambda)]^{-1}e^{ikr}V_{1}(r)Z_{p}(k).
\label{3.43}
\end{equation}Relation \eqref{3.42} follows directly from
\eqref{3.20}, \eqref{3.36} and \eqref{3.43}. $\Box$

We will need the following definition:
\begin{Dn}\label{ Definition 3.10}The potential $V(r)$ belongs to the  Rollnik class if the operator
\begin{equation}A(\lambda)f=\int_{\BR^3}|V(r)|^{1/2}\frac{\exp\{i|k||r-s|\}}{|r-s|}|V(s)|^{1/2}f(s)ds.
\label{3.44}\end{equation} belongs to the Hilbert-Schmidt class.\end{Dn}
\begin{Tm}\label{Theorem 3.11} Let the conditions of Theorem \ref{Theorem 2.5} be fulfilled. Then, for each
$\lambda{\in}E_{+}$  the operators $T_{p}(\lambda)$  belong to the Hilbetrt--Schmidt class.\end{Tm}
\emph{Proof.} Using the equality $Z_{p}(q)Z_{p}^{*}(q)=I_{2}$, we derive relation 
\begin{align}& [\mathcal{F}^{*}_{p}\mathcal{F}_{p}f] \label{3.45}
\\ & \nn =
4{\pi}a(|k|)\int_{\BR^3}V_{1}(r)
\frac{\sin(|k||r-s|)}{|k||r-s|}V_{1}(s)
f(s)ds\label{3.45}\end{align}
for $\mathcal{F}_{p}$ given by \eqref{3.40}.
Since  $V(r)$ satisfies the conditions of Theorem \ref{Theorem 2.5}, the  function $\|V(r)\|$ belongs to the Rollnik class (see \cite[Ch. 1]{Sim}). Now,   it follows 
from \eqref {3.45} that the operators
$\mathcal{F}^{*}_{p}\mathcal{F}_{p}$ belong to the Hilbert--Schmidt class. Using
\eqref{3.42}, we see that operators $T_{1}(\lambda)$ and $T_{2}(\lambda)$ belong to the Hilbert--Schmidt class as well. $\Box$

\section{Scattering amplitude}
In this section, we will investigate  the connection between the stationary scattering problem
(scattering amplitude)   and dynamical scattering problem (scattering operator).

\emph{We suppose further in the text that the conditions of Theorem \ref{Theorem 2.5}  are fulfilled.}
Consider the vector functions
\begin{equation}f_{s,n}(q,k)=\wh {g}_{s}^{*}(q)f(\omega,k,n),\label{4.1}\end{equation}
where $f(\omega,k,n)$ is defined by  \eqref{2.22}.
 Using relations  \eqref{2.22} and  \eqref{3.20}
 we obtain
\begin{equation}
f_{s,n}(q,k)={\nu}T_{s,n}(q,k),\label{4.2}\end{equation}
where
\begin{equation}\nu=2\pi^{2}\lambda.\label{4.3}\end{equation}
According to Theorem \ref{Theorem 3.6}, the wave operators $W_{\pm}(\mathcal{L},\mathcal{L}_{0})$  are complete. Hence, the operators $S_{1}(\lambda)$ and $S_{2}(\lambda)$ given by \eqref{3.39} are unitary in the space $\mathcal{H}$. Then,  there exist  complete orthonormal systems of eigenvectors $G_{j,1}(\omega,\lambda)$ and $G_{j,2}(\omega,\lambda)$  of the operators $S_{1}(\lambda)$ and $S_{2}(\lambda)$, respectively. We denote the corresponding eigenvalues by $\mu_{j,1}(\lambda)$ and  $\mu_{j,2}(\lambda)$. We note that  $|\mu_{j,1}(\lambda)|=|\mu_{j,2}(\lambda)|=1.$ We introduce the vector functions
\begin{equation}F_{s,1}=\col[\overline{f_{s,1}},\overline{f_{s,2}}],\quad (s=1,2); \quad F_{s,2}=\col[\overline{f_{s,3}},\overline{f_{s,4}}],\quad (s=3,4).
 \label{4.4}\end{equation}
  From \eqref{2.22}, \eqref{4.1} and \eqref{4.4}  we have:
 \begin{Pn}\label{Proposition 4.1} The  vector functions $F_{s,p}\,\, (p=1,2)$
 belong to the space  $\mathcal{H}$.\end{Pn}
 Hence,  the vector functions   $F_{s,p}$ can be represented in the form of the series:
 \begin{equation}F_{s,p}(\omega,\omega^{\prime},\lambda)=\sum_{j}G_{j,p}(\omega,\lambda)a_{j}(s,p,\omega^{\prime},\lambda),
 \label{4.5}\end{equation}
where $q=|k|\omega$, $k=|k|\omega^{\prime}$ and
\begin{equation}a_{j}^{*}(s,p,\omega^{\prime},\lambda)=\int_{S^2}F_{s,p}^{*}(\omega,\omega^{\prime},\lambda)G_{j,p}(\omega,\lambda)
d\Omega(\omega). \label{4.6}\end{equation}We represent the $2{\times}1$ vector functions $G_{j,p}(\omega,\lambda)$ in the form  
\begin{align}& G_{j,1}(\omega,\lambda)=\col[G_{j,1}^{(1)}(\omega,\lambda),G_{j,1}^{(2)}(\omega,\lambda)],\label{4.7-}
\\ &
G_{j,2}(\omega,\lambda)=\col[G_{j,2}^{(3)}(\omega,\lambda),G_{j,2}^{(4)}(\omega,\lambda)].\label{4.7}
\end{align}
Relations \eqref{4.4}, \eqref{4.7-} and \eqref{4.7} imply that
\begin{equation}F_{s,1}^{*}G_{j,1}=[T_{s,1}G_{j,1}^{(1)}+T_{s,2}G_{j,1}^{(2)}]\nu,\quad s=1,2,\label{4.8}\end{equation}
\begin{equation}F_{s,2}^{*}G_{j,2}=[T_{s,3}G_{j,2}^{(3)}+T_{s,4}G_{j,2}^{(4)}]\nu,\quad s=3,4.\label{4.9}\end{equation}
The relation
\begin{equation} T_{p}(\lambda)G_{j,p}(\lambda)=-i[\mu_{j,p}(\lambda)-1]G_{j,p}(\lambda) \label{4.10}
\end{equation} is valid.
Using \eqref{3.36}--\eqref{3.39} and  \eqref{4.8}--\eqref{4.10} we obtain:
\begin{equation}a_{j}^{*}(s,p,\omega,\lambda)=-i\nu_{1}[\mu_{j.p}(\lambda)-1]G_{j,p}^{(s)}(\omega,\lambda),\label{4.11}\end{equation}
where
\begin{equation}\nu_{1}=\nu/a(|k|)=2\pi{\sgn}(\lambda)/|k|.   \label{4.12}\end{equation}
Relations \eqref{4.4}, \eqref{4.5} and \eqref{4.11} imply that the following statement is valid.
\begin{Tm}\label{Theorem 4.2} Let the conditions of Theorem \ref{Theorem 2.5}  be fulfilled.
Then we have
\begin{equation}
f_{s,n}(\omega,\omega^{\prime},\lambda)=(\nu_{1}/i)\sum_{j}(\mu_{j,p}(\lambda)-1)G^{(s)}_{j,p}(\omega^{\prime},\lambda)\overline{G_{j,p}^{(n)}(\omega,\lambda)},
\label{4.13}\end{equation}
where  $s=1,2, \,n=1,2,\, \lambda>m,$ when $p=1$ and
$s=3,4,\, n=3,4,\,\lambda<-m,$ when $p=2.$\end{Tm}
\begin{Rk}\label{Remark 4.3} It follows from Theorem \ref{Theorem 3.11} that the series \eqref{4.13} converges in the norm sense.\end{Rk}
Using \eqref{4.1} and \eqref{4.13} we obtain the assertion below.
\begin{Cy}\label{Corollary 4.4 }Let the conditions of Theorem \ref{Theorem 2.5}  be fulfilled.
Then, we have
\begin{equation}
f(\omega,\omega^{\prime},n,\lambda)=(\nu_{1}/i)\sum_{j}(\mu_{j,p}(\lambda)-1)H_{j,p}(\omega^{\prime},\lambda)\overline{G_{j,p}^{(n)}(\omega,\lambda)},
\label{4.14}\end{equation}
where $p=1,\,n=1,2$ if $\lambda>m$; $p=2,\, n=3,4$ if $\lambda<-m$ and
\begin{equation}H_{j,1}=\wh {g}_{1}G_{j,1}^{(1)}+\wh {g}_{2}G_{j,1}^{(2)},\quad
H_{j,2}=\wh {g}_{3}G_{j,2}^{(3)}+\wh {g}_{4}G_{j,2}^{(4)}. \label{4.15}\end{equation}
\end{Cy}
\begin{Rk}\label{Remark 4.5}
Formulas \eqref{4.7-}--\eqref{4.9} describe interconnections between the stationary scattering data  $\{f_{s,p}(\omega,\omega^{\prime},\lambda)\}$ and the dynamical scattering data  $\{G_{j,p}(\omega,\lambda),\,\mu_{j,p}(\lambda)\}$. Thus, formulas \eqref{4.7-}--\eqref{4.9} are quantum mechanical analogs of the ergodic formulas in classical mechanics. 
The quantum mechanical analogs of  ergodic formulas for the radial case $($where $V(r)=V(|r|))$,  were obtained in our papers  \cite{Sakh4, Sakh5}. \end{Rk}
\begin{Rk}\label{Remark 4.6} The analog  of  \eqref{4.13} for the 3D Schr\"{o}dinger equation
was derived in our paper \cite{Sakh6}.\end{Rk}
\section{The scattering by a centre of force}
 In this section, we  study the classical case where
\begin{equation}V(r)=V(|r|),\quad \theta^{\prime}=0,\quad \phi^{\prime}=0.\label{5.1}\end{equation}
Here,  $\theta$ and $\phi$ are angular polar coordinates.
Consider the solution
\begin{equation}\phi(r,k,4,\lambda)=\col[u_1(r,\lambda),u_2(r,\lambda),u_3(r,\lambda),u_4(r,\lambda)]
\label{5.2}\end{equation} 
of the equation \eqref{2.23}. It follows from \eqref{2.9} and \eqref{2.21} that
\begin{align}& u_3(r,\lambda){\sim}e^{ik{\cdot}r}/N+|r|^{-1}e^{i|k||r|}f(\theta,\phi)/N,\quad r{\to}\infty,
 \label{5.3}
 \\ &
 u_4(r,\lambda){\sim}|r|^{-1}e^{i|k||r|}g(\theta,\phi)/N,\quad r{\to}\infty,
 \label{5.4}\end{align} where $N=\norm{g_4}.$
 We use the following assertion
(see \cite[p. 263]{MotMas}).
 \begin{La}\label{Lemma 5.1} Let relations \eqref{5.3} and \eqref{5.4} be valid. The   corresponding complete orthonormal system of eigenfunctions $G_{j,2}(\omega,\lambda)$  of the scattering operators $S_{2}(\lambda),\, \lambda>m$ has the form:
\begin{equation}G_{\ell+1/2}=(4\pi)^{-1/2}[(\ell+1)^{1/2}P_{\ell}(\cos\theta)\wh {\alpha}-
(\ell+1)^{-1/2}P_{\ell}^{1}(\cos\theta)\exp\{i\phi\}\wh {\beta}],\label{5.5}\end{equation}
where $\ell=0,1,2,...$
\begin{equation}G_{\ell-1/2}=-(4\pi)^{-1/2}[\ell^{1/2}P_{\ell}(\cos\theta)\wh {\alpha}+
\ell^{-1/2}P_{\ell}^{1}(\cos\theta)\exp\{i\phi\}\wh {\beta}],\label{5.6}\end{equation}
where  $\ell=1,2,...$\end{La}
Here $P_{\ell}(\cos\theta)$ and $P_{\ell}^{1}(\cos\theta)$ are Legendre polynomials. The matrices $\wh {\alpha}$ and $\wh {\beta}$ are defined by the relations
\begin{equation}\wh {\alpha}:=\col[1,0],\quad
                       \wh {\beta}:=\col[0,1]\label{5.7}\end{equation}
It follows from \eqref{5.5} and \eqref{5.7} that
\begin{align}& G_{\ell+1/2}^{1}=(4\pi)^{-1/2}(\ell+1)^{1/2}P_{\ell}(\cos\theta),
\label{5.8}
\\ &
G_{\ell+1/2}^{2}=-(4\pi)^{-1/2}(\ell+1)^{-1/2}P_{\ell}^{1}(\cos\theta)\exp\{i\phi\}.\label{5.9}\end{align}
Using \eqref{5.6} and \eqref{5.7} we obtain
\begin{align} & G_{\ell-1/2}^{1}=-(4\pi)^{-1/2}\ell^{1/2}P_{\ell}(\cos\theta),
\label{5.10}
\\ &
G_{\ell-1/2}^{2}=-(4\pi)^{-1/2}\ell^{-1/2}P_{\ell}^{1}(\cos\theta)\exp\{i\phi\}.\label{5.11}\end{align}
Taking into account the relations
\begin{equation} P_{\ell}(1)=1,\quad P_{\ell}^{1}(1)=0,
\label{5.12}\end{equation} and using \eqref{4.12}--\eqref{4.15} and \eqref{5.8}--\eqref{5.11} we have
 \begin{align} & f(\theta,\phi)=\frac{1}{2i|k|}\sum_{\ell=0}^{\infty}[\ell+1)(\mu_{\ell+1/2}-1)+\ell(\mu_{\ell-1/2}-1)]P_{\ell}(\cos{\theta}),
\label{5.13}
\\ &
 g(\theta,\phi)=\frac{1}{2i|k|}\sum_{\ell=1}^{\infty}(\mu_{\ell-1/2}-\mu_{\ell+1/2})P_{\ell}^{1}(\cos{\theta})\exp\{i\phi\},
\label{5.14}\end{align}
Recall the classical formulas (see \cite[p. 263]{MotMas}) for the case \eqref{5.1}--\eqref{5.4}:
\begin{align} & f(\theta,\phi)=\frac{1}{2i|k|}\sum_{\ell=0}^{\infty}[\ell+1)(\mathcal{S}_{\ell+1/2}-1)+\ell(\mathcal{S}_{\ell-1/2}-1)]P_{\ell}(\cos{\theta}),
\label{5.15}
\\ &
 g(\theta,\phi)=\frac{1}{2i|k|}\sum_{\ell=1}^{\infty}(\mathcal{S}_{\ell-1/2}-\mathcal{S}_{\ell+1/2})P_{\ell}^{1}(\cos{\theta})\exp\{i\phi\},
\label{5.16}\end{align}
where $\mathcal{S}_{\nu}$  is connected with the phase shifts $\delta_{\nu}$ by the
relation
\begin{equation} \mathcal{S}_{\nu}=\exp\{\delta_{\nu}\}.\label{5.17}\end{equation}
Comparing formulas \eqref{5.13}, \eqref{5.14} and \eqref{5.15}, \eqref{5.16} we
obtain the next assertion.
\begin{Cy}\label{Corollary 5.2} Let the conditions of Theorem \ref{Theorem 2.5} and condition \eqref{5.1} be fulfilled. Then, we see that: \\
1. The following equality is valid
\begin{equation}\mu_{\nu}(\lambda)=\mathcal{S}_{\nu}(\lambda). \label{5.18}\end{equation}
2. The classical results \eqref{5.15}, \eqref{5.16}  may be considered as a special case of the formula \eqref{4.14} .
\end{Cy}
In \cite{Sakh5}, we proved
\eqref{5.18} by the methods of theory of ordinary differential equations.

\begin{flushright}
Lev Sakhnovich,\\
99 Cove ave. Milford, CT, 06461, USA,\\
e-mail: {\tt lsakhnovich@gmail.com}

\end{flushright}

\end{document}